\documentclass[twocolumn,prl,amsmath,amssymb,showpacs,superscriptaddress]{revtex4-1}
\usepackage{epsf}      
\usepackage{graphicx}
\usepackage{color}
\usepackage{soul}
\usepackage{gensymb}
\usepackage{float}
\begin{document}
%\mbox{

\title {Na--ion diffusion and electrochemical performance of NaVO$_3$ anode in Li/Na batteries}

\author{Mahesh Chandra}
\affiliation{Department of Physics, Indian Institute of Technology Delhi, Hauz Khas, New Delhi-110016, India}
\author{Tuhin S. Khan}
\affiliation{Department of Chemical Engineering, Indian Institute of Technology Delhi, Hauz Khas, New Delhi-110016, India}
\author{Rishabh Shukla}
\affiliation{Department of Physics, Indian Institute of Technology Delhi, Hauz Khas, New Delhi-110016, India}
\author{Salahuddin Ahmed}
\affiliation{Department of Mechanical Engineering, Indian Institute of Technology Delhi, Hauz Khas, New Delhi-110016, India}
\author{Amit Gupta}
\affiliation{Department of Mechanical Engineering, Indian Institute of Technology Delhi, Hauz Khas, New Delhi-110016, India}
\author{S. Basu}
\affiliation{Department of Chemical Engineering, Indian Institute of Technology Delhi, Hauz Khas, New Delhi-110016, India}
\author{M. Ali Haider}
\affiliation{Department of Chemical Engineering, Indian Institute of Technology Delhi, Hauz Khas, New Delhi-110016, India}
\author{R. S. Dhaka}
\email{rsdhaka@physics.iitd.ac.in}
\affiliation{Department of Physics, Indian Institute of Technology Delhi, Hauz Khas, New Delhi-110016, India}

\date{\today}                                         

\begin{abstract}
We study Na ion diffusion and electrochemical performance of NaVO$_3$ (NVO) as anode material in Li/Na--ion batteries with the specific capacity of $\approx$350 mAhg$^{-1}$ at the current density 11~mAg$^{-1}$ after 300 cycles. Remarkably, the capacity retains $\ge$200~mAhg$^{-1}$ even after 400~cycles at 44~mAg$^{-1}$ with Coulombic efficiency $>$99\%. The deduced diffusion coefficient from galvanostatic intermittent titration technique (GITT), electrochemical impedance spectroscopy (EIS) and cyclic voltammetry (CV) measurements for NVO as anode in Li--ion battery is in the range of 10$^{-10}$--10$^{-12}$~cm$^2$s$^{-1}$. In case of Na-ion batteries, the NVO electrode exhibits initial capacity of 385~mAhg$^{-1}$ at 7~mAg$^{-1}$ current rate, but the capacity degradation is relatively faster in subsequent cycles. We find the diffusion coefficient of NVO--Na cells similar to that of NVO--Li. On the other hand, our charge discharge measurements suggest that the overall performance of NVO anode is better in Li--ion battery than Na-ion. Moreover, we use the density functional theory to simulate the energetics of Na vacancy formation in the bulk of the NVO structure, which is found to be 0.88~eV higher than that of the most stable (100) surface. Thus, the Na ion incorporation at the surface of the electrode material is more facile compared to the bulk. 
\end{abstract}

\maketitle

\section{\noindent ~Introduction}

In recent years, energy storage devices especially rechargeable batteries have become an integral part of portable electronic devices like mobiles, laptops as well being used at large scale in electrical vehicles (EV), hybrid electrical vehicles and power grids \cite{DunnSci11,TarasconNat01}. Among various types of batteries, in the present scenario the only commercially established and feasible technology is based on Li-ion intercalation, mainly due to the advantage of high energy density, good rate kinetics and light weight  \cite{MaromJMC11,GoodenoughJACS13}. However, the major concern is due to the limited availability and uneven distribution of Li around the world, which evoke the urge among the researchers to look for other possibilities \cite{WangESM16,EftekhariJPS18}. Therefore, even after nearly three decades since first commercial battery was introduced, constant efforts are going on to search new electrode materials (negative as well as positive) for improving its electrochemistry \cite{KimAEM12,QiJMCA17,EftekhariESM17,JulienLBST16}. Among the alternative avenue, the Na-ion batteries are one of the potential candidates particularly for grid level storage due to its low cost and abundance on the globe \cite{PuSmall19,JameshJPS18,HanEES15,LiuMT16}, and therefore extensive research has been started in this field over past few years \cite{PalomaresEES12,WangNC15,RuiAM15,KangJMCA15,BarpandaNC14,MaheshMRB18,MaheshCI18}. In this direction, one of the major challenges is to find the suitable electrode materials and in particular a stable anode with high electrochemical performance \cite{MiguelAEM17}. For the Li-ion batteries, though the capacity is limited by the formation of solid electrolyte interface (SEI), graphite has been used as a commercial anode material owing to its low redox potential against lithium (Li/Li$^+$), good electronic conductivity, high Li-ion mobility (10$^{-8}$-10$^{-10}$~cm$^2$s$^{-1}$) and great structural stability during the process of Li intercalation and deintercalation  \cite{FlandroisCarbon99}. However, in general the same may not be suitable for Na-ion batteries as the interlayer distance of graphite ($\approx$0.34~nm) is insufficient to accommodate the larger Na$^+$--ion \cite{CaoNL12} and the Na-intercalated graphite compounds are thermodynamically not stable \cite{MoriwakeRSCad17,LiuPNAS16,StevensJES01}. Therefore, in most of the studies, Na metal is being used as counter electrode (anode) for testing the suitability of cathode materials in half cell configuration. In this context, more recently Conder {\it et al.} and  Pfeifer {\it et al.} reported very important results on the reactivity of Na metal with most commonly used few electrolytes and found high interfacial resistance and significant degradation in carbonate-based electrolytes \cite{ConderCC19, PfeiferCPS19}. 

Due to these reasons it is vital to search for stable anode materials, and transition metal oxides (TMOs) have been considered as an alternative electrodes in Li--/Na--ion battery technology because of their higher reversible capacities along with safer operating potentials \cite{GoripartiJPS14,SuESM16}. Among the TMOs, the vanadium based oxides have been extensively studied as anode and cathode for both Li and Na-ion batteries, where various oxidation state of vanadium ranging from +5 to +2 have been utilized \cite{JianJPS13,PralongCM12,SongJPS10,LiuJMC09,SarohaACSO19}. More recently, the NaVO$_3$ (theoretical capacity about 220 mAhg$^{-1}$ with $n=$ 1) has been investigated as  anode material in Li-ion batteries \cite{LiuJSSE16} as well as both cathode and anode in the Na-ion batteries  \cite{VenkateshEC14,AliACSAMI18,ZhangJPS18}. The only report on NaVO$_3$ (NVO) as anode in Li-ion batteries manifest a higher capacity of 623~mAhg$^{-1}$ in the initial cycle, but the capacity degraded by more than 50\% after few cycles \cite{LiuJSSE16}. In a Na-ion battery, an initial discharge capacity of 359~mAhg$^{-1}$ is reported for NVO as an anode, but the capacity retain nearly one third of initial capacity after 80 cycles \cite{AliACSAMI18}. Whereas, as a cathode in a Na-ion battery, an initial capacity of 120~mAhg$^{-1}$ has been achieved by activating the NVO higher than 4.5~V, where vanadium does not changes its 5+ valence state during charge/discharge \cite{ZhangJPS18}. Here, an anionic redox reaction has been suggested to be responsible for the source of electrons during electrochemical reactions \cite{ZhangJPS18}. Also, the morphology and particle size of the electrode material play a crucial role in the overall performance especially the specific capacity and cycle life of a battery \cite{ZhangJPS18,RoyJMCA15,WangNanoscale10}. The main reason for this is the large surface to volume ratio and shorter diffusion length in nanosized materials. However, the compromise is with the capacity degradation upon cycling due to undesirable surface reactions and other side reactions associated with large surface area and surface energy \cite{RoyJMCA15,WangNanoscale10}. The other disadvantages are low volumetric energy density (important for EV applications) and the difficult procedures associated with synthesizing naomaterials \cite{WangNanoscale10}.

Here, it is important to note that, to the best of our knowledge the NaVO$_3$ is largely unexplored as an anode material in both Li--/Na--ion batteries \cite{LiuJSSE16,AliACSAMI18}. Also, detailed study with long cycling and systematic comparizon of the diffusion coefficient using different methods are still missing. Therefore, in the present work we have synthesized NaVO$_3$ by conventional solid state reaction method and show that although the initial capacities for the NVO electrode materials synthesized by sol gel method (i.e. small particle size) \cite{LiuJSSE16} in a Li-ion battery are higher, the cycle life is far better for the electrodes with large particle size. The capacity of 200~mAhg$^{-1}$ for a high current density of 110~mAg$^{-1}$ is retained after 100~cycles. We have deployed various electrochemical measurements, including galvanostatic intermittent titration technique (GITT), which is most reliable measurement to obtain diffusion coefficient, to get insight of the critical parameters for the performance of a battery. Our study reveals better electrochemical performance (in particular cycle life) of NVO anode for both Li--/Na--ion batteries. Also, we observed using density functional theory that the Na ion incorporation at the surface is more facile compared to the bulk of the electrode material. 
 
\section{\noindent ~Experimental details}

The NaVO$_3$ (NVO) sample was synthesized by solid-state reaction method using precursors V$_2$O$_5$ and Na$_2$CO$_3$  as vanadium and sodium sources, respectively. The stoichiometric amount of the precursors were mixed and ground using agate mortar for 5--6 hrs to get a homogeneous mixture. The ground mixture was calcined at 600$^o$C for 12 hrs to obtain fine powder of NVO. The crystal structure of the synthesized sample was analyzed by x-ray diffraction (XRD) using a Cu-K$\alpha$ radiation ($\lambda$ = 1.5406~\AA). We used scanning electron microscopy (SEM) to check the morphology and get an idea about the particle size of the NVO powder sample. The x-ray absorption spectroscopy (XAS) measurements at vanadium K-edge have been performed in the transmission mode at the energy scanning beamline (BL-09) \cite{BasuJPCS14} of the INDUS-2 synchrotron centre at RRCAT, India.

To fabricate the electrode, the slurry was prepared by mixing active material NVO (80\%) with polyvinylidene fluoride (PVDF, 10\%) as a binder and super-P carbon black (10\%) as conductive agent dissolved in N-methyl pyrrolidinone (NMP) solvent. The slurry was coated on the copper (thickness of 15 $\mu$m) foil by using film coater and doctor blade. Subsequently, the coated sheet was dried in the oven at 80$^o$C for 12~hrs to remove any traces of the solvent. The thickness of the dried sheet was reduced by 10\% of the original thickness by calendaring, to minimize the pores generated due to the solvent evaporation. Then, circular disks of diameter 14~mm and 19~mm were punched from the calendared electrodes and separator (Glass fiber filter GB-100R), respectively, which were finally used to assemble the coin cells. The average weight of the active material in an electrode is 3.3~mg. The 2016 type coin-cells were assembled using a crimping machine (MTI Corp.) inside a nitrogen filled glove box (Jacomex). The electrolyte used was 1M LiPF$_6$ (for Li-ion based coin cell) dissolved as the solute in 1:1 ethylene carbonate (EC) and dimethyl carbonate (DMC) and NaClO$_4$ (for Na-ion based coin cell) dissolved as the solute in 1:1 EC and DEC. The coin cells were assembled in half-cell (working electrode against lithium disk and sodium disk) format to investigate electrochemical performance of NVO electrodes. A potentiostat (Palmsens) was used for the cyclic voltammetry (CV) and battery cycler (Bio-Logic-VMP3) for electrochemical impedance spectroscopy (EIS) and GITT measurements. The galvanostatic charge discharge data were collected using a cycler from Neware (BTS400).
 
 \section{\noindent ~Computational Method}
 
Periodic plane-wave based density functional theory (DFT) method viena ab-initio simulation package (VASP) \cite{KresseCMS96} was applied to calculate the Na vacancy formation energy at the bulk and surface of NVO anode material. Bulk geometry of NVO was adapted from theâ `Materials Project' database \cite{JainAPLM13} materials ID mp-19083 \cite{PerssonUS14}. Na is modelled as body centered cubic (BCC) bulk structure with lattice parameter, $a=$ 4.2906~\AA~ and $\alpha=$ 90$^o$. Energy and force convergence criteria of 1$\times$10$^{-4}$~eV and 0.05~eV\AA$^{-1}$ are set respectively for geometry optimization calculations. Revised Perdew-Burke-Ernzerhof \cite{HammerPRB99} GGA exchange correlation functional (RPBE) is used along with projector augmented wave (PAW) potentials \cite{KressePRB99} and the plane wave energy cut-off value of 396~eV. To calculate the Na vacancy formation energy at NVO bulk, a 3$\times$3 supercell is used to minimize the interaction between two neighbouring Na vacancies. All of the calculations  performed under spin-polarized condition, with Hilbert U$_{eff}$ parameter of value 3.25~eV employed for V atom. In order to study the nano-structuring effect; three different low-index surfaces, (100), (110) and (111) are modelled. A super cell of size 2$\times$1 used for (100) and (110) surfaces, whereas (111) surface was modelled with 1$\times$1 unit cell. All the three surfaces modelled using a four layer surface slab; wherein two bottom layers kept fixed to their bulk positions and the top two layers allowed to relax. A vacuum of size 20~\AA applied above the surface of the slab. The calculation of Na vacancy formation energy performed by removing single Na atom from the surface.

The surface formation energy (E$_f$) of the three terminations, (100), (110) and (111) have been calculated using the formula $$E_f = \frac{E_{surface}-E_{bulk}}{2A}$$ where E$_{surface}$ and E$_{bulk}$ are the energies of NVO surface and bulk, respectively and A denotes the area of the corresponding surface. The Na vacancy formation energy (E$_v$) is calculated using the formula below;$$ E_v= E(Na_{x-1}VO) + E(Na)- E(Na_xVO) $$ 
where  E(Na$_x$VO), E(Na$_{x-1}$VO) and E(Na) are the energies of NVO structure E(Na$_{x-1}$VO), E(Na$_x$VO) formed after Na removal and energy of bulk Na, respectively. For the NVO bulk Monkhorst--pack \cite{MonkhorstPRB76} k-points grid of 1$\times$3$\times$2 utilized, whereas k-point grid of 2$\times$1$\times$1 is used for the (100), (110) and (111) surfaces.  Monkhorst--pack k-point grid of 5$\times$5$\times$5 is used for Na bulk lattice geometry optimization.

\section{\noindent ~Results and discussion}

\begin{figure}[h]
\includegraphics[width=3.3in]{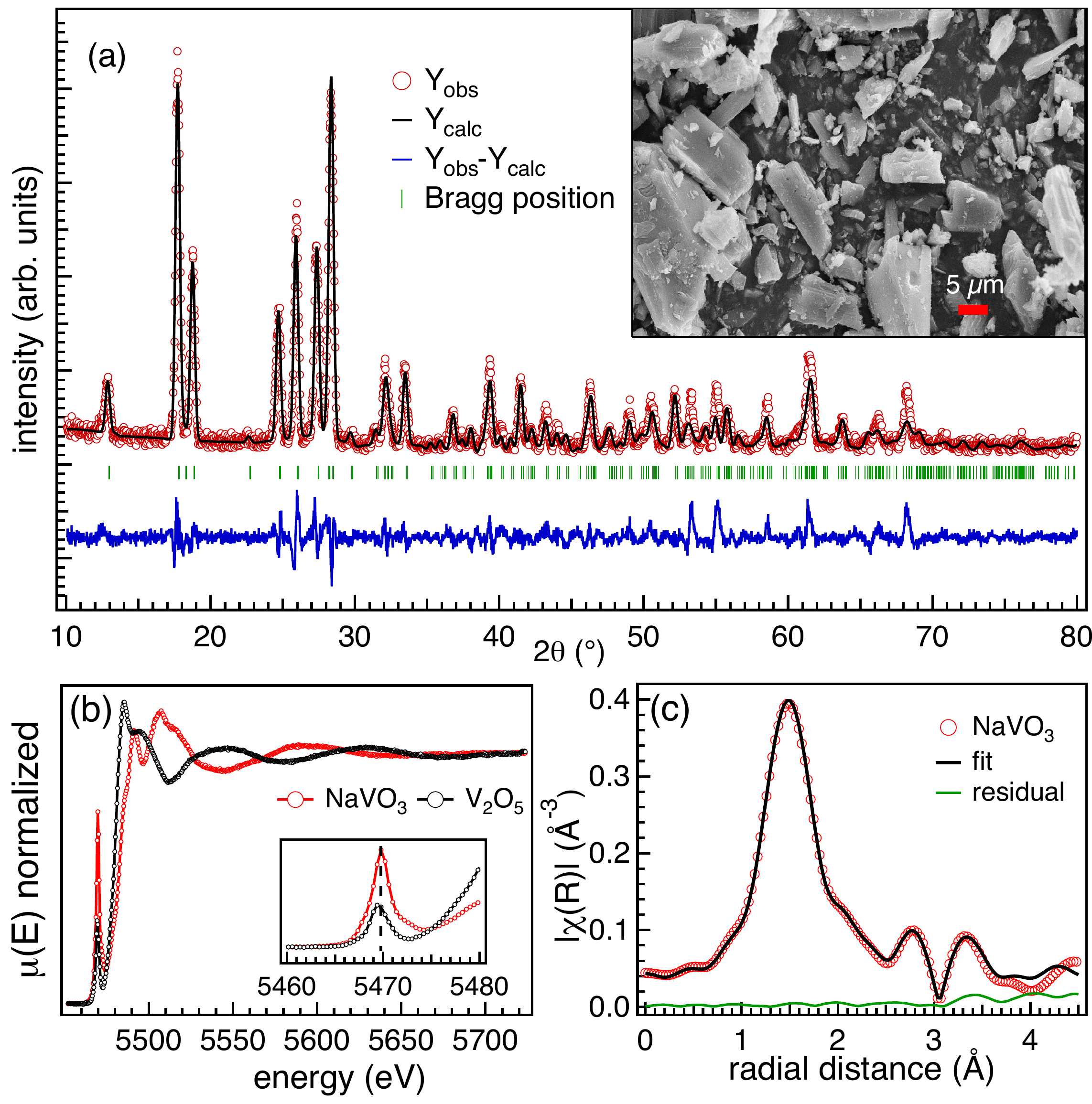}
\renewcommand{\figurename}{Figure}
\caption {(a) The XRD pattern along with the Rietveld refinement using the monoclinic space group C1c1, the inset shows SEM image of the synthesized NaVO$_3$ powder sample. (b) the normalized x-ray absorption spectra of NaVO$_3$ along with the reference V$_2$O$_5$ sample, inset is the zoomed view of the pre-edge feature, and (c) the fitted $\chi$(R) vs radial distance (R) spectra considering only single scattering path.}
\label{fig:XRD}
\end{figure}

The phase purity and crystal structure of NaVO$_3$ sample have been determined by the powder XRD measurements performed at room temperature, as shown in the Figure~\ref{fig:XRD}(a). The Rietveld refinement of the XRD pattern has been done using fullprof software and the extracted lattice parameters are: $a=$ 10.543~\normalfont\AA, $b=$ 9.458~\normalfont\AA , $c=$ 5.876~\normalfont\AA, and $\beta=$ 108.42\normalfont$^0$ with the goodness of fitting ($\chi^2$) and weighted profile parameter (R$_{wp}$) as 2.24 and 26.3, respectively. The structure is monoclinic with C1c1 space group, which is in agreement with the earlier reports \cite{VenkateshEC14,AliACSAMI18}. The ratio of Na and V ions has been deduced from the energy-dispersive x-ray measurement, which found to be Na:V as 1:0.98. 
The inset of Figure~\ref{fig:XRD}(a) shows the SEM image of the synthesized powder. The morphology is of irregular shape with particle size in micrometers. Further, in order to confirm the local structure, we have performed the x-ray absorption spectroscopy (XAS) measurements in the transmission mode. Figure~\ref{fig:XRD}(b) presents the normalized near edge x-ray absorption (XANES) spectra along with the extended x-ray absorption spectrum (EXAFS) region for the NaVO$_3$ sample and V$_2$O$_5$ (considered as reference sample), measured at the vanadium K-edge. We observed the pre-edge peak position [see inset in Figure~\ref{fig:XRD}(b)] at 5469.5~eV for both the samples, which confirms the presence of vanadium in 5+ oxidation state in our sample. On the other hand, the strong intensity of the pre-edge feature for the NaVO$_3$ sample is due to the coordination of vanadium ions in the tetrahedral symmetry of monoclinic phase and having the d$^0$ configuration, in agreement with previous reports \cite{AliACSAMI18,LeeCM16}. The fitted EXAFS spectrum ($\chi(R)$ versus $R$ plot) of NaVO$_3$ sample is shown in Figure~\ref{fig:XRD}(c) where we used the structural parameters obtained from the XRD analysis and considering only the single scattering paths from the vanadium atom, which has the two sites with the same symmetry and different atomic positions. The main peak in $\chi(R)$ versus $R$ plot is observed at about 1.5~\AA~, which is reported to be
corresponds to the interaction between nearest vanadium and oxygen atoms \cite{AliACSAMI18}. In addition, the vanadium atom has two neighboring sodium atoms with the different interatomic distances, as the corresponding peaks are observed between 2 and 3.5~\AA~ in Figure~\ref{fig:XRD}(c) for the pristine NaVO$_3$ sample. The electrochemical properties of the NVO powder has been studied as an active electrode material against Li (NVO--Li) as wells as against Na (NVO--Na) as counter electrodes in half cell configurations. In the upcoming sections, we will discuss the electrochemical properties and performance of the NVO--Li and NVO--Na cells followed by the discussion on the results from density functional theory. 

\begin{figure}
\includegraphics[width=3.25in]{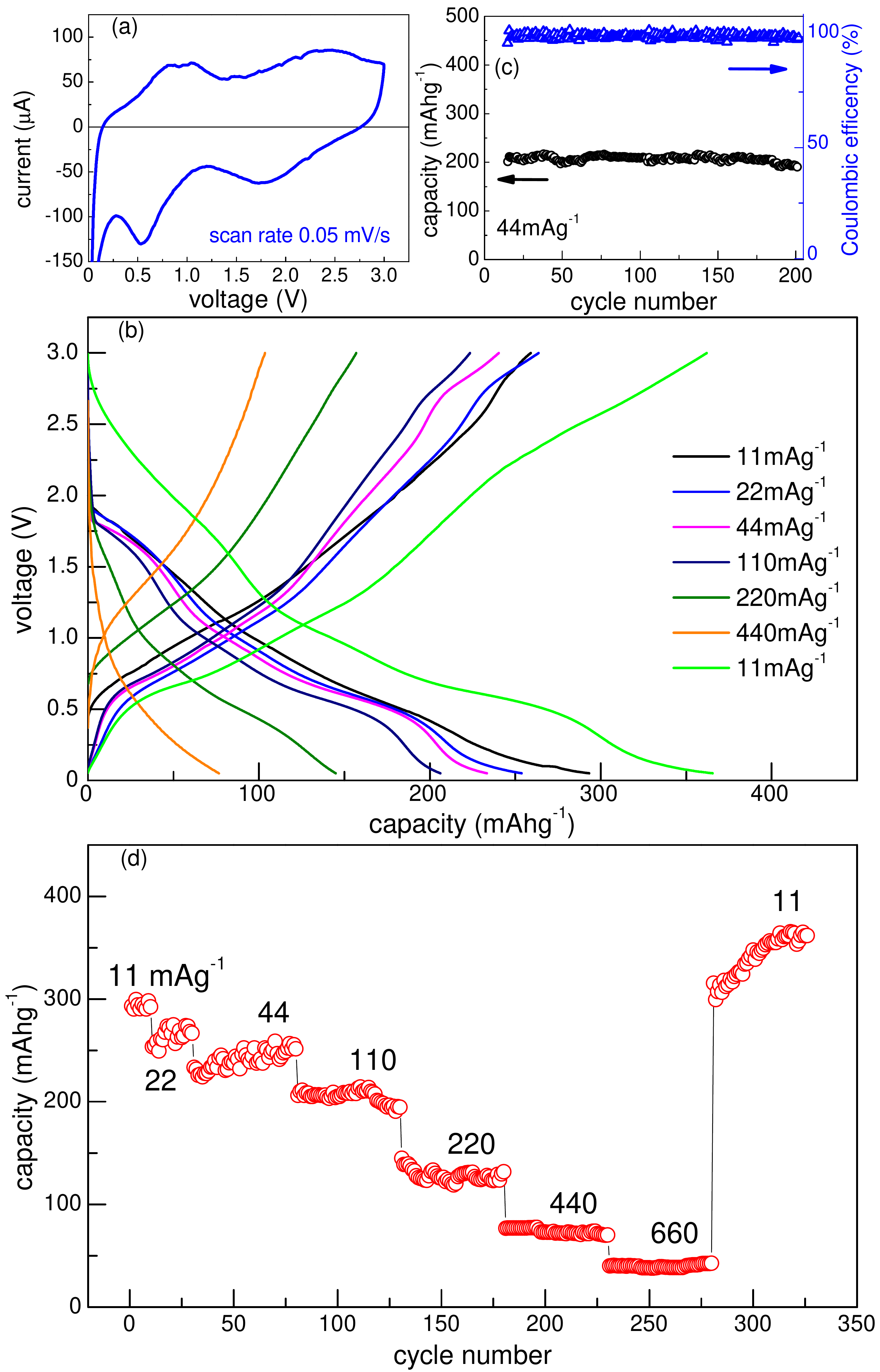}
\renewcommand{\figurename}{Figure}
\caption {For NVO--Li cells: (a) CV measurement performed at 0.05~mVs$^{-1}$ scan rate, (b) the galvanostatic charging discharging cycles at different current densities, (c) the capacity variation measured at 44~mAg$^{-1}$ current density with increasing cycle number after 200~cycles at lower current rates along with the Coulombic efficiency on right scale, (d) specific capacity versus cycle numbers at different current densities.}
\label{fig:Fig2_LiCD}
\end{figure}

In order to get insight of the electrochemical properties we first perform cyclic voltametry on the NVO--Li half cells, as shown in the Figure \ref{fig:Fig2_LiCD}(a). A slow scan rate (0.05~mVs$^{-1}$) was applied to obtain distinct redox peaks within the voltage range from 0.01~V to 3~V in the CV measurements. Two redox couples can be seen in the CV plot, one at lower potentials (0.5~V and 0.8~V) and other at higher potentials (1.8~V and 2.3~V), which indicate a reversible electrochemical reaction in the electrode material. We have further performed the galvanostatic charging discharging at various current densities, as shown in Figure~\ref{fig:Fig2_LiCD}(b). The capacity performance with increasing cycle numbers for current density ranging from from 11~mAg$^{-1}$  to 660~mAg$^{-1}$ is shown in Figure~\ref{fig:Fig2_LiCD}(d). We have achieved a capacity of 300~mAhg$^{-1}$ at 11~mAg$^{-1}$ current rate in the initial cycle and the capacity remains higher than this even after 300~cycles at different current rates and when finally come back to 11~mAg$^{-1}$, see Figure~\ref{fig:Fig2_LiCD}(d). We note here that the capacity performance in this study is far more stable with cycling than the previously reported for nanosized NaVO$_3$ \cite{LiuJSSE16} prepared using sol gel method where the capacity decreases from 356~mAhg$^{-1}$ to nearly 250~mAhg$^{-1}$ just after 30~cycles at much lower current rate of 5~mAg$^{-1}$. Moreover, if we compare the results of microsized electrode in ref \cite{LiuJSSE16}, there the capacity degradation is faster and it reduces from 250~mAhg$^{-1}$ to 100~mAhg$^{-1}$ after only 30~cycles. Whereas, in the present case, despite of the high current density used [see for example capacity at 110 mAg$^{-1}$ current density in Figure~\ref{fig:Fig2_LiCD}(d)], the capacity is more than 200~mAhg$^{-1}$ after nearly 100~cycles at various current rates. The specific capacity decreases below 100~mAhg$^{-1}$ only when we increase current rate beyond 220~mAg$^{-1}$. Interestingly, after a total 285~cycles at different current rates, the measured capacity at current density of 11~mAg$^{-1}$, is higher than that in the first cycle of a fresh cell and it is quite steady ($\ge$350~mAhg$^{-1}$) from 300 to 330~cycles, as presented in the Figures~\ref{fig:Fig2_LiCD}(b) and (d). In Figure \ref{fig:Fig2_LiCD}(c), we have shown the long cycling results from another NVO--Li cell measured at 44~mAg$^{-1}$ current density, which was earlier tested with different current rates for 200~cycles (all current rate results are not show to avoid repetition). The capacity as wells as Coulombic efficiency are very stable even after 200~cycles (effectively 400~cycles) at the current density of 44~mAg$^{-1}$.

\begin{figure*}
\includegraphics[width=7.1in]{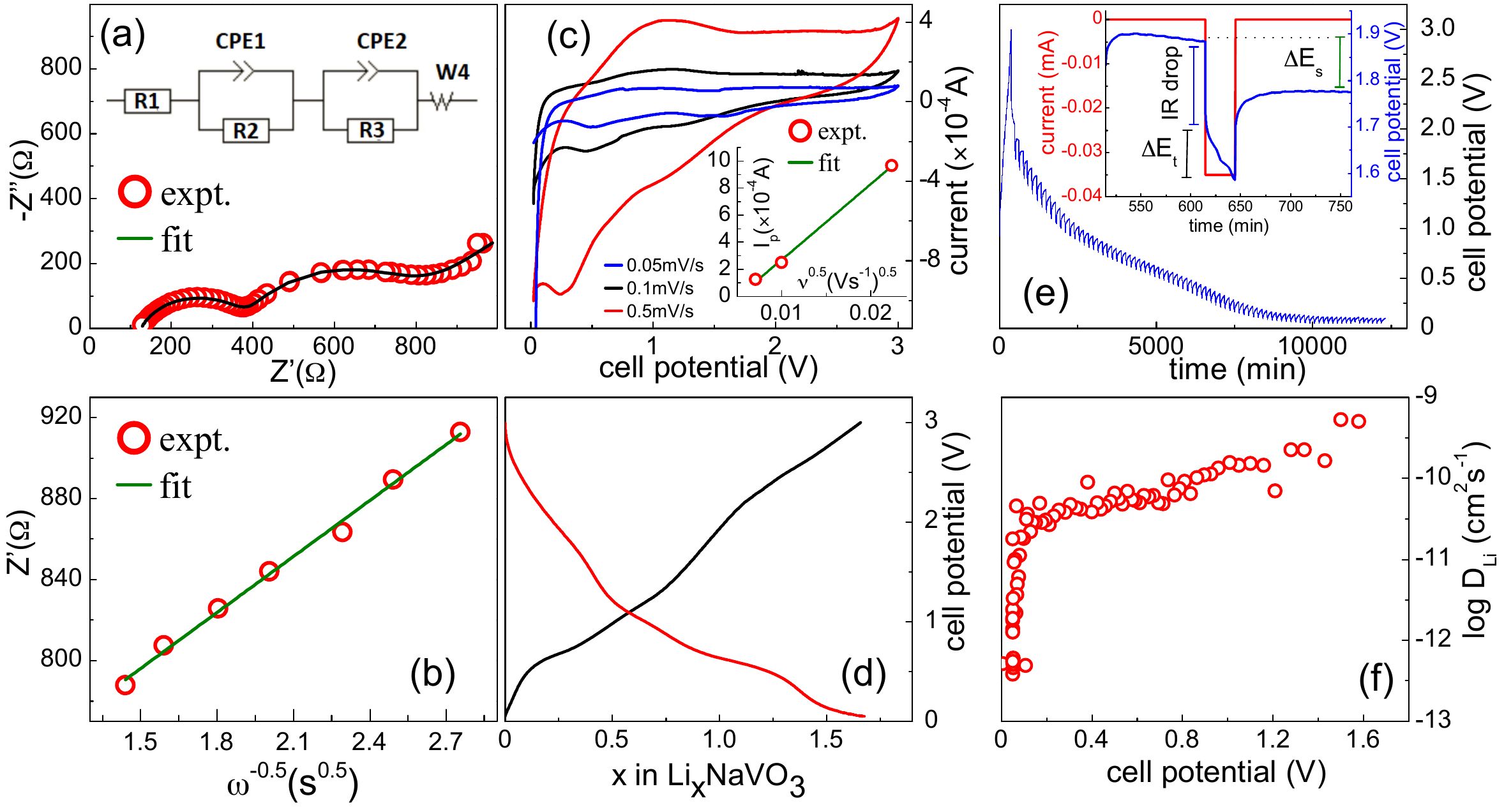}
\renewcommand{\figurename}{Figure}
\caption {For NVO--Li cells: (a) EIS plot with fitting and its equivalent circuit in the inset, (b) linear fit of Z' with $\omega^{-0.5}$ in the low frequency range, (c) CV measurement performed at different scan rates and the inset shows the variation of anodic peak current with square root of the scan rate, (d) the variation of Li concentration during charge discharge calculated from the 300$^{th}$ cycle at slow current density of 11~mAg$^{-1}$, (e) Galvanostatic intermittent tritation technique (GITT) at 11~mAg$^{-1}$ current rate, (f) the diffusion coefficient calculated from the GITT measurement at different potentials.}
\label{fig:Fig3_LiDC}
\end{figure*}

The diffusion rate or characteristic time ($\tau$) for diffusion of Li/Na ion in an electrode mainly depends upon two important parameters that are the diffusion length (L), which is related to the morphology/particle size and the diffusion coefficient (D), which is the material's property. These quantities are correlated as: $\tau$=L$^2$/D \cite{RoyJMCA15}. For the nanosized particles the diffusion length is small, which may leads to a fast intercalation and hence higher capacity. On the other hand, the bulk particles have advantages in terms of volumetric density and cycling life as mentioned in the introduction as well \cite{RoyJMCA15,WangNanoscale10}. Here, we have investigated the diffusion coefficient of Li ions in the NaVO$_3$ electrode material in a NVO--Li half cell using three different methods namely EIS, CV and GITT. The EIS measurement of the NVO--Li half cell has been performed (after one cycle CV) at 0.9~V in a frequency range of 10~mHz to 10~kHz, as the plot is shown in the Figure~\ref{fig:Fig3_LiDC}(a). The linear variation of the real part of the impedance with inverse of the square root of the applied frequency (in low frequency range) is shown in the Figure \ref{fig:Fig3_LiDC}(b). The impedance measured in EIS is a combination of different contributions from the working electrode, electrolyte and counter electrode. The EIS plot consists of two semicircles at high frequency and medium frequency followed by a straight line at low frequencies. The equivalent circuit for the EIS data is shown in the inset of Figure~\ref{fig:Fig3_LiDC}(a), where R1 represents the bulk resistance of the cell which includes the resistance due to electrodes, separator and electrolyte \cite{ZhangEA06}. As the semicircles are not perfect which could be a manifestation of roughness of the electrode surface, we have used the constant phase element (CPE) instead of normal capacitance. Here, CPE1 and R2 are the capacitance and resistance of SEI layers, respectively. The second semicircle at moderate frequency range represents the charge transfer resistance (R3) and double layer capacitance (CPE2) at the electrode-electrolyte interface. Finally, the straight line at low frequencies is the Warberg impedance which is directly associated with the diffusion of metal ion at the electrode-electrolyte interface. Using this part of the EIS data one can calculate the diffusion coefficient D using the following relation: 
\begin{eqnarray}
	D= \frac{1}{2}\left[\frac{V_M}{FA\sigma}\right]\left[\frac{dE}{dx}\right]^2
	\end{eqnarray}
where, V$_M$ is the molar mass of the active material, A is area of the electrode, dE/d$x$ is the slop of cell voltage to the Li ion concentration which is obtained from the plot in Figure \ref{fig:Fig3_LiDC}(d). F is the Faraday constant (96486~Cmol$^{-1}$) and $\sigma$ is the Warburg impedance coefficient which is obtained by the slope of linear fit between Z' and $\omega^{-0.5}$ [Figure~\ref{fig:Fig3_LiDC}(b)]. The relation between the real part of impedance and frequency $\omega$ is as follow
\begin{eqnarray}
	 Z'= R_s +R_{ct} +{\sigma\omega^{-0.5}}
	 \end{eqnarray}
where, R$_s$ and R$_{ct}$ are the ohmic resistance and charge transfer resistance, respectively. The deduced value of $\sigma$ from Figure \ref{fig:Fig3_LiDC}(b) is 92.10~\ohm/s$^{0.5}$ and substituting this value in the equation (1) the diffusion coefficient comes out to be 8.1$\times$10$^{-12}$~cm$^2$s$^{-1}$.

In cyclic voltammetry (CV) measurements, the anodic/cathodic peak current values depend upon the scan rates as clearly seen in Figure~\ref{fig:Fig3_LiDC}(c) and it is also related to the diffusion coefficient by the following relation:
\begin{eqnarray}
 I_p= 2.69 \times 10^5 ACD^{1/2}\nu^{1/2}n^{3/2}
  \end{eqnarray}
where, I$_p$ is the peak current, A is area of the electrode, $\nu$ is the scan rate, C is the bulk concentration of Na equal to 0.0238~mole/cm$^3$ in present case, and $n$ is the number of electrons that are reversibly involved in the redox reaction extracted from the Figure~\ref{fig:Fig3_LiDC}(d). This plot is extracted from the 300th charging-discharging cycle of NVO--Li cell at the very slow current rate of 11~mAg$^{-1}$, where the capacity is quite steady. The maximum number of electrons ($n$) involved in the reaction from this plot are found to be $n=$ 1.6 at the lowest measured potential. The CV for NVO--Li half cell at different scan rate has been performed and plotted in Figure~\ref{fig:Fig3_LiDC}(c). The absolute value of anodic peak current (I$_p$) at different scan rates versus the square root of scan rate ($\nu^{0.5}$) is plotted. From this plot in the inset of Figure~\ref{fig:Fig3_LiDC}(c), by taking the slope of the linear fit and using equation (3), we deduced the value of diffusion coefficient, which found to be 1.35$\times$10$^{-11}$~cm$^2$s$^{-1}$.
 
  \begin{figure}[h]
\includegraphics[width=3.4in]{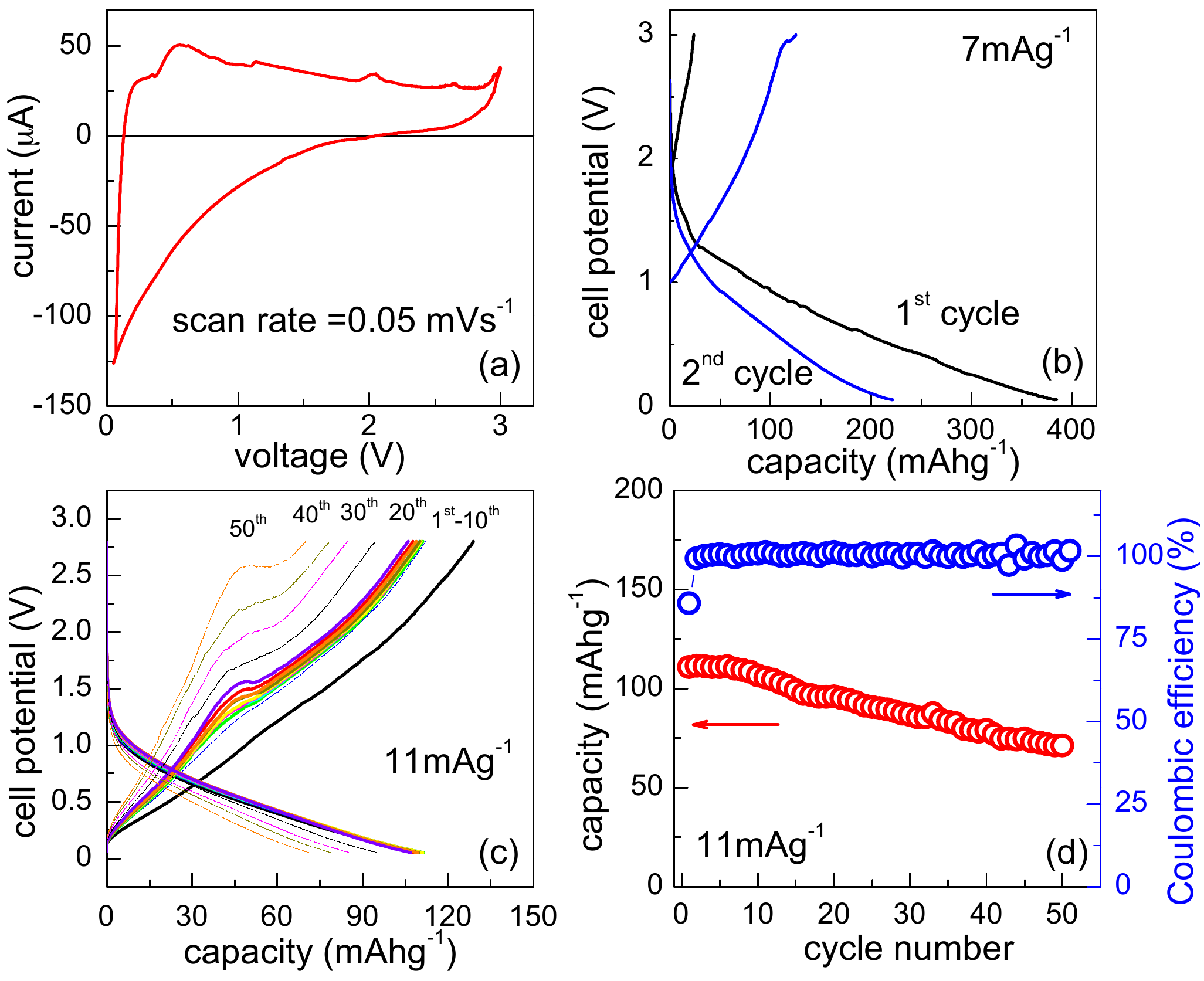}
\renewcommand{\figurename}{Figure}
\caption {For NVO--Na cells: (a) CV measurement performed at 0.05~mVs$^{-1}$ scan rate, (b) the charging discharging cycle profile for first two cycles measured at 7~mAg$^{-1}$ current density, (c)  the charging discharging profile for 50~cycles at 11~mAg$^{-1}$ current density, (d) specific capacity and Coulombic efficiency change with cycle numbers at 11~mAg$^{-1}$.} 
\label{fig:Fig4_NaCD}
\end{figure}

 In order to further understand the kinetics of Li ion insertion/deinsertion in the NVO--Li half cell, the GITT measurements have been performed, which considered as a standard and more reliable method to deduce the diffusion coefficient. In GITT measurement a positive/negative (during charging/discharging) current pulse is applied to the electrode, which is interrupted by a rest time where there is no current applied. During the rest time, the Na ions diffuse into the electrode material to compensate the stoichiometric gradient in the electrode. We have used 0.07~mA current (corresponding to the current density 22~mAg$^{-1}$) to charge the electrode up to 3~V and GITT is performed during discharge with a small current 0.035~mA (corresponding to the current density 11~mAg$^{-1}$). The duration of the current pulse was 30~mins and the rest time of 100~mins. The GITT curve form 3~V to 0.05~V is shown in Figure~\ref{fig:Fig3_LiDC}(e). The diffusion coefficient is given by the following equation for the charging/discharging current being sufficiently small:
	\begin{eqnarray}
	D= \frac{4}{\pi \tau}\left[\frac{m_B V_M}{M_B A}\right]^2\left[\frac{\Delta E_s}{\Delta E_t}\right]^2
	\end{eqnarray}
here $\tau$ is the duration of the current pulse in sec, M$_B$  (g mol$^{-1}$) is the molecular weight of the electrode material, V$_M$ is the molar volume, m$_B$ is the weight of the active material in gram which is 0.0033~g in present case, and A is the electrode surface area. $\Delta$ E$_s$ denote the change in the steady-state voltage at a plateau and $\Delta$ E$_t$ is the total change in cell voltage during pulse time as shown in the inset of Figure~\ref{fig:Fig3_LiDC}(e). These two voltages are measured for each step to calculate the diffusion coefficient for the respective step. The diffusion coefficient calculated for the first step is 8.9$\times$10$^{-10}$~cm$^2$s$^{-1}$, which is considerably swift. Such fast kinetics of Li-ion diffusion in NVO is possibly responsible for the significantly large capacity and good cycling life in NVO--Li half cell. An increasing trend in the log of diffusion coefficient with different steps and corresponding cell potential is shown in Figure~\ref{fig:Fig3_LiDC}(f). 

 \begin{figure}[h]
\includegraphics[width=3.5in]{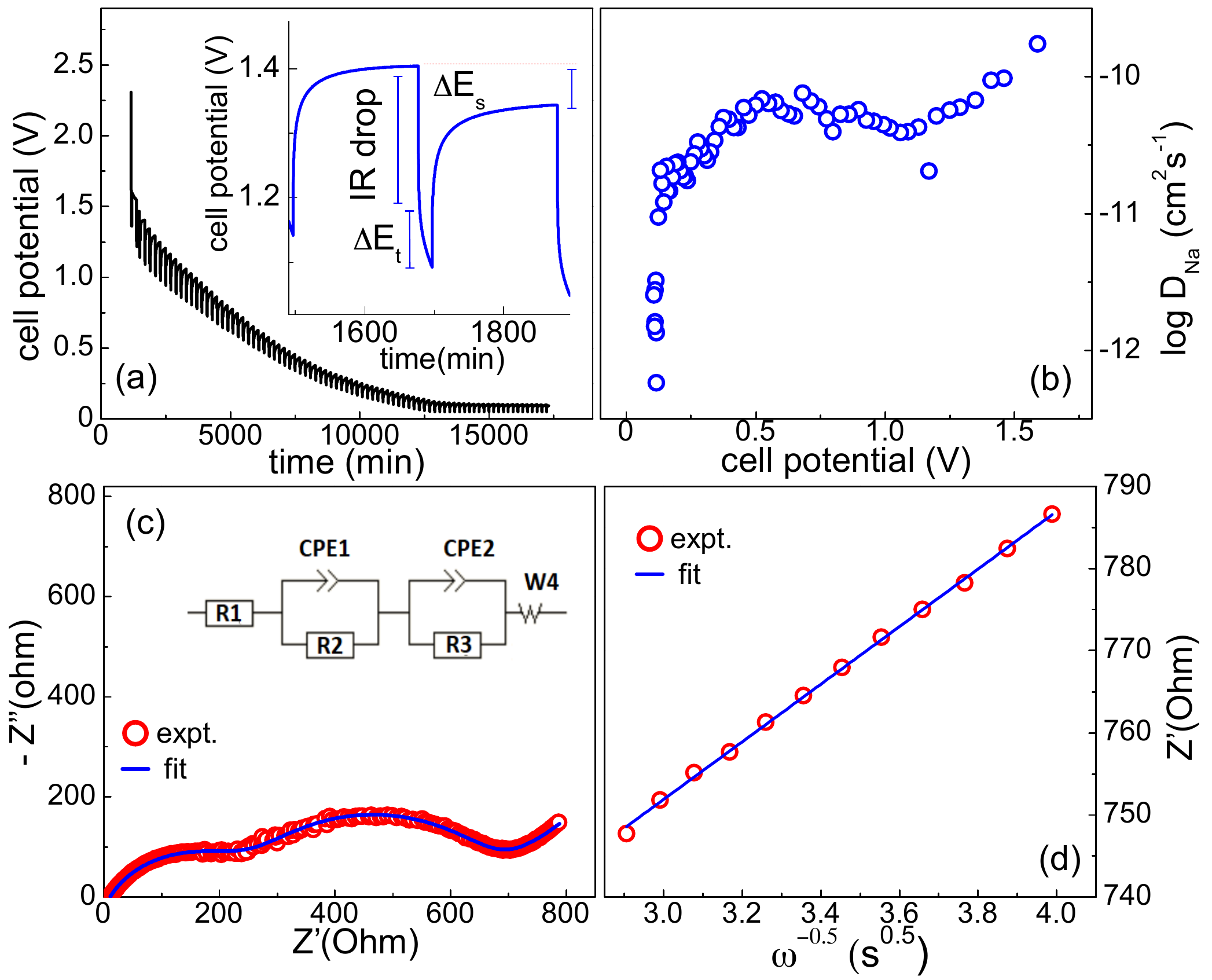}
\renewcommand{\figurename}{Figure}
\caption {For NVO--Na cells: (a) the GITT measurement for cell discharge at 11~mAg$^{-1}$ current density, (b) the deduced diffusion coefficient variation with cell potential, (c) the EIS plot with fitting and its equivalent circuit in the inset, and (d) linear fit of Z' with $\omega^{-0.5}$ in the low frequency range.} 
\label{fig:Fig5_NaDC}
\end{figure}

Now we discuss the electrochemical performance of NVO as an anode in a Na--ion battery and therefore, we have performed electrochemical  measurements on the half cell with NVO as anode and Na as counter electrode. The CV and charging discharging of a NVO--Na half cell have been shown in Figure~\ref{fig:Fig4_NaCD}(a--c). The CV performed at 0.05~mVs$^{-1}$ scan rate shows redox peaks below 0.5~V. The initial capacity for NVO--Na cell is as high as 385~mAhg$^{-1}$ at 7~mAg$^{-1}$ current density but it reduces quite rapidly in the next cycle as shown in the Figure~\ref{fig:Fig4_NaCD}(b). We have shown the 50~cycles of charging-discharging at 11~mAg$^{-1}$ current density in Figure~\ref{fig:Fig4_NaCD}(c) and the variation of discharge capacity and Coulombic efficiency with cycling in Figure~\ref{fig:Fig4_NaCD}(d). The initial capacity at 11~mAg$^{-1}$ is 110~mAhg$^{-1}$ and capacity retention is nearly 65~$\%$ after 50~cycles. On the other hand, note that the Coulombic efficiency remains almost 100~$\%$ up to 50~cycles except for the first cycle. The low Coulombic efficiency is possibly due to SEI layer formation in the first cycle. It is clear that there is a lot of scope for improving the performance of NVO--Na batteries. One way is increasing the conductivity between highly insulating NVO matrix by increasing the additive (such as acetylene black) as done by Ali {\it et al.} \cite{AliACSAMI18}. The authors achieved higher capacity, but its retention upon cycling is still needs further investigation. Another way could be the carbon coating on active material, which increase the surface contact or modification in the structure by appropriate substitution at V site and controlling Na content.

\begin{figure}
\includegraphics[width=3.3in]{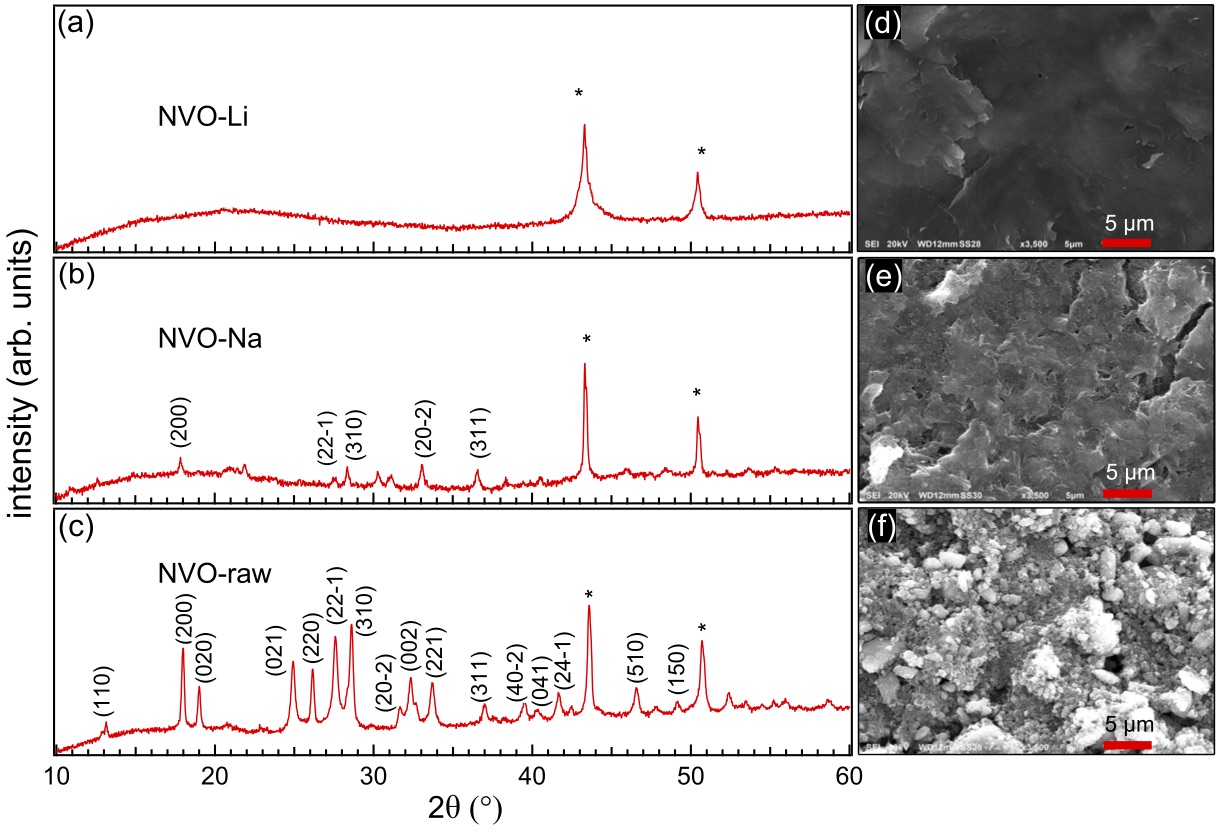}
\renewcommand{\figurename}{Figure}
\caption {The ex-situ XRD patterns and corresponding SEM images of cycled electrodes for NVO--Li cell (a, d) after 325~cycles, and (b, e) for NVO--Na cell after 60~cycles, (c, f) the XRD pattern and corresponding SEM image of the unused NVO electrode. The star(*) in the XRD patterns  indicates the peaks from Cu foil used as current collector.} 
\label{fig:Fig6_Exsitu}
\end{figure}

 \begin{figure}
\includegraphics[width=3.47in]{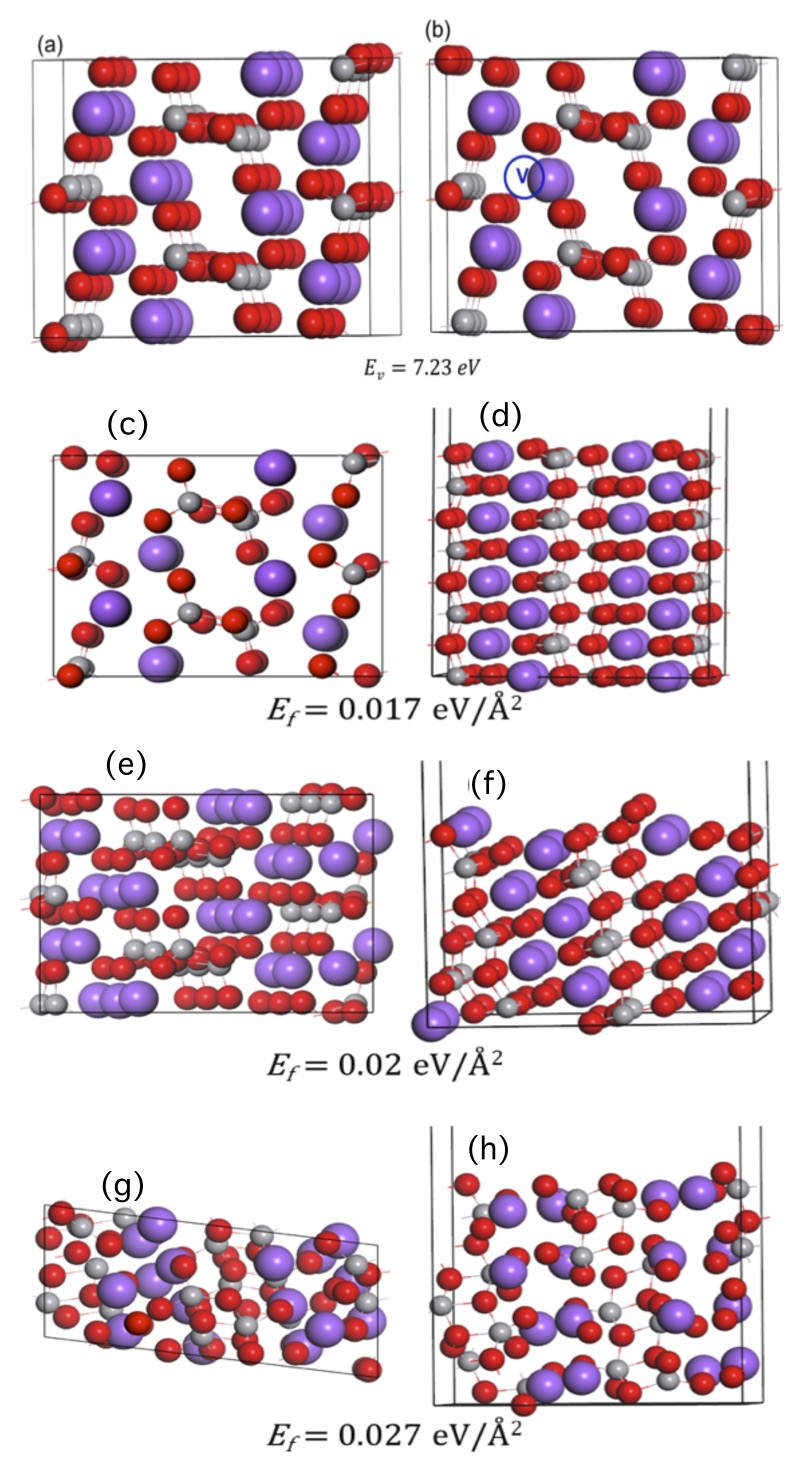}
\renewcommand{\figurename}{Figure}
\caption {The DFT optimized geometry of NaVO$_3$ (a) bulk and (b) single Na vacancy in bulk. The position of Na vacancy shown as blue circle with V sign and corresponding Na vacancy formation energy (E$_v$) are given below each panel. The optimized geometry for different surfaces (b, c) 100, (d, e) 110 and (f, g) 111 surfaces. Top and side views are shown in (c, e, g) and (d, f, h) panels and the corresponding surface formation energy (E$_f$) are given below.} 
\label{fig7}
\end{figure}

 \begin{figure*}
\includegraphics[width=7.0in]{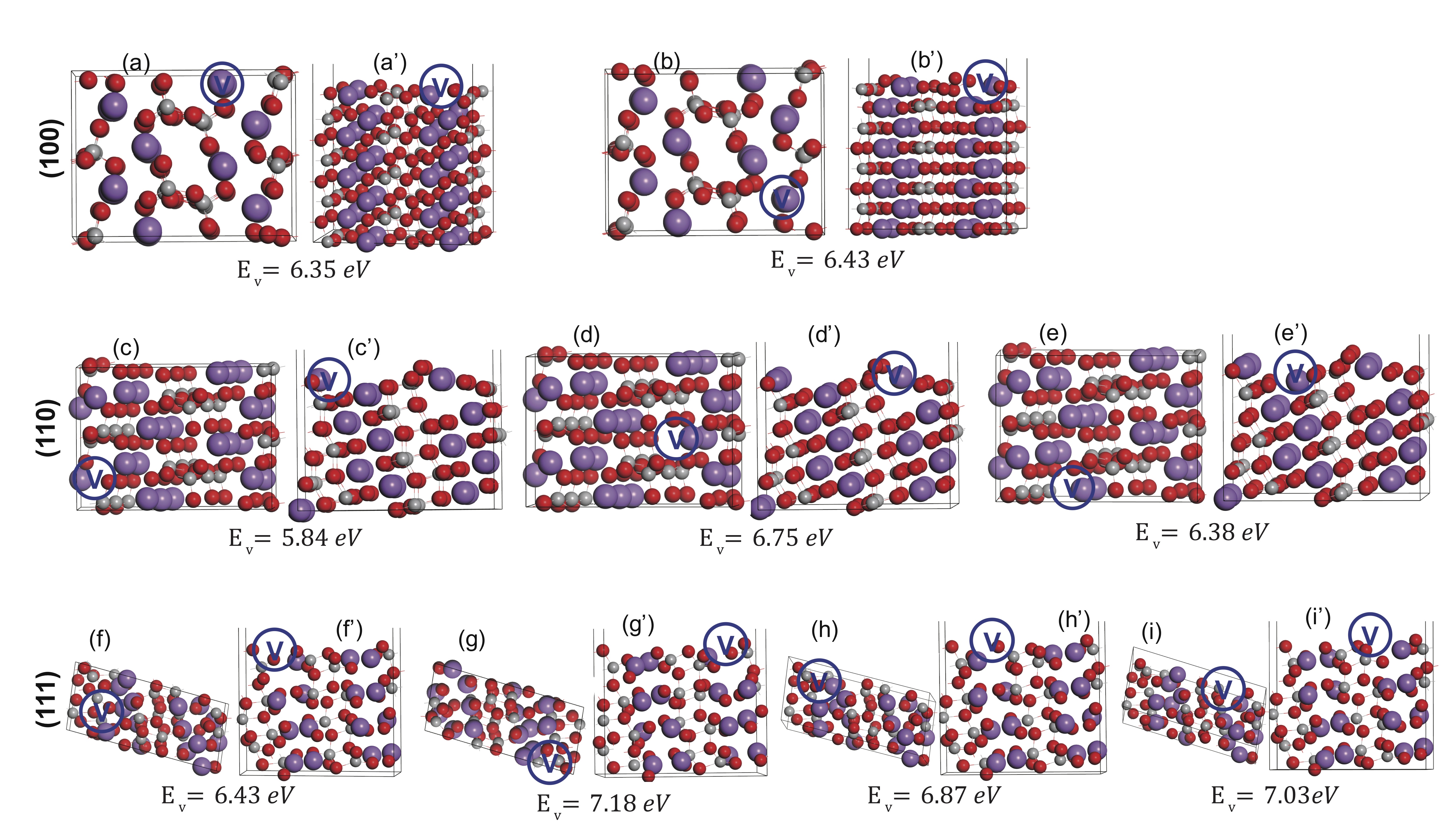}
\renewcommand{\figurename}{Figure}
\caption {The DFT optimized geometry of distinct vacancy configuration at the NaVO$_3$ surface;  (a, a') and (b, b') for (100) surface, (c, c'), (d, d') and (e, e') for (110) surface and (f, f'), (g, g'), (h, h') and (i, i') for the (111) surface, respectively, showing top view (left figure) and side view (right figure) along with the corresponding Na vacancy formation energy (E$_v$) given below.} 
\label{fig8}
\end{figure*}

Moreover, in order to get the insight about the diffusion of Na ions in NVO anode for NVO--Na cell, after one cycle CV, we have performed the GITT and EIS (at 0.3~V in a frequency range of 10~mHz to 200~kHz) measurements similar to that for NVO--Li cell. The results are shown in Figure~\ref{fig:Fig5_NaDC}(a--d), in a similar fashion as Figure~3. For the GITT, we have used a 0.07~mA current (corresponding to 22~mAg$^{-1}$ current density) to charge the electrode up to 3~V and measurement has been performed during discharge with a small current 0.035~mA. The duration of the current pulse was 30~mins with the rest time of 120~mins. Following the same procedure, we found the diffusion coefficient of 1.8$\times$10$^{-10}$~cm$^2$s$^{-1}$ for the first GITT step and 2.55$\times$10$^{-12}$~cm$^2$s$^{-1}$ at 0.1~V. The variation of diffusion coefficient at different voltage is shown in the Figure~\ref{fig:Fig5_NaDC}(b), which indicates that the diffusion become sluggish at very low voltages. The EIS curve and its equivalent circuit is shown in Figure~\ref{fig:Fig5_NaDC}(c). The linear fit of the real part of the impedance with inverse square root of the frequency are shown in the Figure~\ref{fig:Fig5_NaDC}(d). Further, we determined the diffusion coefficient calculated using equations (1) and (2), which come out 3.65$\times$10$^{-11}$~cm$^2$s$^{-1}$. Interestingly, the diffusion coefficient of Na ions in NVO is comparable with that of Li ions, but capacity degradation for NVO--Na half cell is high possibly due to the large size of Na ion compare to that of Li ion. Also, we note there that the Na diffusion coefficient and consequently the electrochemical performance is much improved as compare to the ref.~\cite{AliACSAMI18}.

After the electrochemical cycling measurements, we have performed the ex-situ XRD measurements and SEM imaging on the cycled electrodes of NVO--Li as well as NVO--Na cells to understand the crystal structure and morphology of intercalated electrode  material after SEI layer formation. We compare the results with respect to those of the un-cycled NVO electrode coated on the Cu foil. We have opened the cells after discharging them to 0.1~V. The results of ex-situ XRD and their respective SEM are shown in Figure~\ref{fig:Fig6_Exsitu}(a--f). The XRD and SEM image of unused coated electrode is shown in Figures~\ref{fig:Fig6_Exsitu}(c, f) respectively, which confirm all the reflections similar to the powder sample with irregular particle size in the electrode. The ex-situ XRD of NVO--Li cell in Figure~\ref{fig:Fig6_Exsitu}(a) shows amorphous behavior and only peaks corresponding to Cu current collector (shown by *) appears in the XRD. This is consistent with the earlier report \cite{LiuJSSE16}. Similarly the ex-situ XRD pattern of the NVO--Na [Figure~\ref{fig:Fig6_Exsitu}(b)] cell shows the presence of some reflections from crystallographic planes indicating that the crystal structure of the electrode material is not completely amorphous after Na insertion and repeated cycling. The corresponding SEM images [Figures~\ref{fig:Fig6_Exsitu}(d) and (e)] for both cycled electrodes show a clear SEI layer with agglomeration of the particles to form patches. 

Finally, in order to understand the link between the electrode performance and energetics of Na vacancy creation in the structure of the NVO electrode, Na vacancy formation energy in NVO bulk and surfaces calculated using the DFT method. The optimized structures of NVO bulk is shown in Figure~7(a). For the NVO bulk, Na vacancy formation energy calculated to be high, E$_v^{Na}$ =7.23~eV. The corresponding structure is shown in Figure~7(b). To obtain the Na vacancy formation energy at the surface, three different low index surfacse (100), (110) and (111) constructed, as shown in Figure~7(c)-(d), (e)-(f) and (g)-(h) panels, respectively. The (100) surface of NVO found to be the most stable with a surface energy 0.017~eV\normalfont\AA$^{-2}$, followed by (110) surface having surface energy 0.020~eV\AA$^{-2}$. The (111) surface termination of NVO found to be least stable with surface energy 0.027~eV\AA$^{-2}$. Na vacancy is created on the surface by removing a single Na atom. Two distinct Na vacancy configuration at the (100) surface are shown in Figure~8(a, a') and (b, b'). The Na vacancy formation energy are similar for the two configurations, E$_v$ for configuration (a) and (b) calculated to be 6.35 and 6.43~eV, respectively, which is approximately 0.88~eV lower compared to the Na vacancy formation energy in NVO bulk. Similarly Na vacancy formation energy is calculated for three distinct vacancy configuration at the (110) surface, as shown in Figure~8(c, c'), (d, d') and (e, e'). The Na vacancy formation is most favourable for the configuration (c, c') with E$_v$=5.84~eV which is nearly 0.5~eV lower compared to the vacancy formation energy at the (100) surface. The Na vacancy formation energy at the (110) surface for configuration (d) and (e) are less favourable with E$_v$=6.75 and 6.38~eV, respectively. The Na vacancy formation energy at the (111) surface found to be less favourable, with E$_v$ calculated to be 6.99, 7.18, 6.87 and 7.03~eV for configuration Figure~8 (f, f'), (g, g'), (h, h') and (i, i'), respectively. From the DFT calculated energies it can be concluded that the Na vacancy formation energy is greatly reduced at the (100) and (110) surfaces compared to the NVO bulk, with reduction as much as 0.88 and 1.39~eV for the two surfaces, respectively. This assert to the fact that on applying a suitable synthesis method, the propensity to form surface Na vacancies is greatly enhanced, thereby showing increased capacity (356 mAhg$^{-1}$) in experiments for nanosized NVO \cite{LiuJSSE16}. However, on prolong operation the stability of the surface structure remains a concern to maintain the capacity. This is further reflected in significant drop in the performance of the nanoscale material on operation in charging--discharging mode \cite{LiuJSSE16}, while in the present case the capacity of NVO (solid-state synthesized) anode remained steady at 300 mAhg$^{-1}$ after the long cyclic operations at different current rates.

\section{\noindent ~Conclusions}

In summary, we have synthesized NaVO$_3$ by the solid state reaction method, and investigated as an anode material for Li as well as Na ion batteries. The NVO--Li and NVO--Na half cells have been studied for their electrochemical properties. The specific capacity of 350~mAhg$^{-1}$ at 11~mAg$^{-1}$ current density has been obtained after nearly 300~cycles for NVO--Li half cell. The capacity is very stable as compared to the previously reported microsized electrode synthesized using sol gel method and a capacity of $\ge$100~mAhg$^{-1}$ is obtained at high current density of 220~mAg$^{-1}$. Hence, we can clearly see the bulk NVO anode shows superior electrochemical performance than nanosized NVO in a Li--ion battery. The deduced diffusion coefficient for NVO--Li is first time reported using the GITT technique. The diffusion coefficient from GITT and EIS measurement is 5.05$\times$10$^{-10}$~cm$^2$s$^{-1}$ (for the first GITT step) and 8.1$\times$10$^{-12}$~cm$^2$s$^{-1}$ respectively. The diffusion coefficient for 5.45$\times$10$^{-13}$~cm$^2$s$^{-1}$ for the last step of the GITT measurement indicating that the Li kinetics is sluggish at low potentials. For NVO--Na half cell the specific capacity of nearly 385~mAhg$^{-1}$ has been obtained at 7~mAg$^{-1}$ for the first cycle and a relatively fast degradation nearly 40~$\%$ capacity is observed in the next cycle. The diffusion coefficient using GITT is 1.8$\times$10$^{-10}$~cm$^2$s$^{-1}$ at higher potentials and that using EIS technique is 3.65$\times$10$^{-11}$~cm$^2$s$^{-1}$. Our study reveal that the NVO is a promising anode material for Li--ion batteries and for Na--ion batteries it needs further investigation with for example carbon coating or substitution with other appropriate metal ion to improve the cycle life. The DFT calculated energetics for Na vacancy creation on the surface and in the bulk of the NVO structure is pointing towards the effect of applied synthesis method resulting into differently formed surface structures with varying surface to volume ratio, which invariantly translates into significantly improved capacity. Thus, adaptation of a synthesis method to generate stable structures having more surface vacancies, holds the key to improve electrochemical performance. 

\section{\noindent ~Acknowledgments}

MC and RS thank SERB-DST (NPDF, no PDF/2016/ 003565), and DST-Inspire, respectively, for the fellowship. RSD acknowledges the financial support from SERB-DST through Early Career Research (ECR) Award (project reference no. ECR/2015/000159). We thank the IIT Delhi for providing central research facilities (XRD, EDX, and SEM) for characterization and HPC facility for computational resources. RSD and SB acknowledge the financial support from IIT Delhi through the FIRP project (IRD no. MI01418). RS and RSD thank Ravi Kumar and S. N. Jha for help and support during XAS measurements at RRCAT, India. The authors also acknowledge the financial support from DST through ``DST--IIT Delhi Energy Storage Platform on Batteries" project (no. DST/TMD/MECSP/2K17/07).

\end{document}